\documentclass[twocolumn,showpacs,preprintnumbers,amsmath,amssymb]{revtex4}
\usepackage[dvips]{graphicx}

\begin{document}
\title{A proposal for sympathetically cooling neutral molecules
using cold ions}
\author{F. Robicheaux}
\affiliation{Department of Physics, Purdue University, West Lafayette,
Indiana 47907, USA}
\date{\today}

\begin{abstract}
We describe a method for cooling neutral molecules that have
magnetic and electric dipole moments using collisions with cold
ions. An external magnetic field is used to
split the ground rovibrational energy levels of the molecule.
The highest energy state within the ground rovibrational manifold
increases in energy as the
distance to the ion decreases leading to a repelling potential.
At low energy, inelastic collisions are strongly suppressed due to
the large distance of closest approach. Thus, a collision between
a neutral molecule and a cold ion will lead to a decrease in the
molecule's kinetic energy with no change in internal energy.
We present results for
the specific case of OH molecules cooled by Be$^+$, Mg$^+$, or Ca$^+$ ions.
\end{abstract}

\pacs{37.10.Mn, 34.50.Cx, 37.10.Pq}

\maketitle

For over 10 years, there has been a substantial experimental effort
to cool molecules to $\mu$K temperatures. This interest
is sparked by the possibility of collective effects in a cold, dense
molecular gas,\cite{GSL,MBZ} interesting collisional mechanisms at low
energy,\cite{RVK,PZH} or to enhance spectroscopic accuracy needed for precision
measurements.\cite{EAH,HKS} A variety of techniques have been explored.
Reference~\cite{WDG} used buffer gas cooling to trap CaH at a
temperature of $\sim 400$~mK; by a specific choice of scattering
conditions, a single Ar-NO collision produced NO molecules at
$\sim 400$~mK.\cite{EVC}
The Stark effect has been used to slow and trap a variety of
molecules with electric dipole moments.\cite{MBM,CWB}
Reference~\cite{SBD} took advantage
of the favorable Franck-Condon factors in SrF to perform one-dimensional
laser cooling.
Evaporative cooling was able to decrease the temperature of trapped OH from an initial
45~mK to 5.1~mK.\cite{SHY}
References~\cite{ZMP,FR1} proposed variations of a Sisyphus cooling
where each photon removes a large fraction of translational energy
of the molecule; the method in Ref.~\cite{ZMP} was realized in
Ref.~\cite{ZEG} to cool CH$_3$F from 390~mK to 29~mK.
Reference~\cite{ELP} was able to remove 95\%
of the translational energy of an O$_2$ beam using a ``molecular
coilgun".
Mechanical effects, as in a spinning nozzle\cite{MGH}, can
produce colder molecules by having the molecules exit in a
moving frame of reference.
This is not a complete list, but the experimental limit for cooling
preexisting molecules is still above 1~mK 10 years after Ref.~\cite{DFK}.
There are also a wide variety of theoretical proposals for
cooling molecules into the ultracold regime; some examples are
in Refs.~\cite{TRC,BTB,TSZ,WLZ,LVH,ICZ,AWH}.

The purpose of this paper is to
describe a method to cool molecules well below
1~mK using standard cooling techniques already present in many labs
while losing only a small fraction of the molecules.
The basic idea is to use the collision between a neutral
molecule and an ultracold positive ion for cooling
the translational motion of the molecule. The specific situation
we examine is the case where the molecule has both a magnetic and
an electric dipole moment with the molecule in a specific internal
state. The specific internal state can be achieved through the
trapping process, the natural cooling near the nozzle of a molecular
beam, or active cooling as in Ref.~\cite{MHL}.
We assume the collision takes place in
a uniform B-field; for molecules held in a magnetic trap,
the B-field varies in space but can be considered to be uniform
over distance scales that characterize the scattering with the
ion. The E-field from the ion at the molecule leads to a
repulsive $1/r^4$ interaction while the splitting of the energy levels
due to the B-field leads to a ``high" frequency scale which
allows the collision to be adiabatic.

The essential difference between the current proposal and
sympathetic cooling from atoms (e.g.~Ref.~\cite{LBP}) is that
the collisions take place at longer range which vastly increases
the elastic collision rate while decreasing the rate for
changing the internal state of the molecule.
Our method is most similar to the sympathetic cooling of
molecular ions by laser cooled atomic ions (e.g.~Ref.~\cite{TWW})
except the long range repulsive $1/r$ interaction is
replaced with a repulsive $1/r^4$ interaction;
it is well known that sympathetic cooling leads to translationally
cold molecular ions while leaving the internal state unchanged.
All of our calculations are performed for OH molecules
but we expect the basic cooling mechanism to work for other
molecules as well.

\begin{figure}
\resizebox{80mm}{!}{\includegraphics{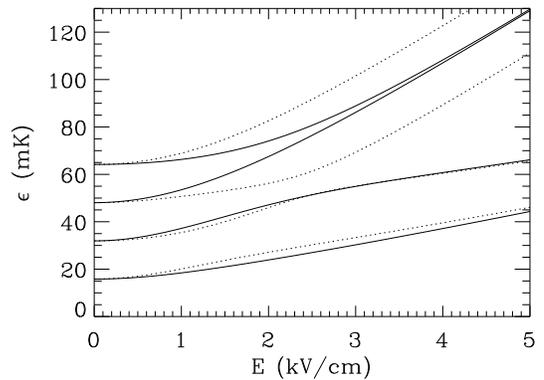}}
\caption{The 4 largest energies of OH in a B-field
of 300~G as a function of applied E-field. The solid lines are
for an angle of 90$^\circ$ between the fields; the
dotted lines are for 45$^\circ$.}
\end{figure}

Figure~1 shows how the 4 highest internal energies of OH vary
with E-field strength when a 300~G B-field is
also applied. The two different line types are for different
angles between the fields. If the OH
starts in the uppermost state, an increasing electric field
causes the internal energy to increase. In a collision, the
E-field from the ion at the OH increases as the
distance decreases which leads to an increasing internal
energy. This increasing internal
energy is equivalent to a potential energy that increases with
decreasing separation, i.e. a repelling force between
the ion and the neutral molecule which keeps them separated.
If the angle between the fields is not too close to 90$^\circ$,
the collision will be adiabatic and the OH will remain in the
uppermost state. If there is a close approach {\it and} the
angle is near 90$^\circ$, the near degeneracy between the 
upper two levels could allow a transition between the states.
Fortunately, if the OH and the ion have small relative kinetic
energy, they will not be able to approach close enough for
a transition between the states to occur. For a 300~G
B-field, if the OH kinetic energy is less than 20~mK,
then the inelastic cross section is $10^5$ times smaller than
the elastic cross section. The ratio of collision rates for
different B-fields are shown in Fig.~3.

Our calculations were performed using a mixture of classical and
quantum mechanics. The relative motion of the ion and
molecule is treated classically. This should be a good approximation
over the energy range presented here. We discuss the limitations
of our calculation below and argue that the classical rates should
be a good approximation down to $\mu$K temperatures.
For the internal states
of the molecule, we solved the full time dependent Schr\"odinger
equation using the leapfrog algorithm. By solving the Schr\"odinger
equation for this 8-state system, we ensure that no approximation
causes the inelastic processes to be erroneously suppressed.

For the B-field strengths considered in this paper, the OH
molecule has 8 internal states of interest: 4 spin states times 2
parity states.
The Hamiltonian we use is the explicit form in Ref.~\cite{SYS} Appendix
A.2 but modified to include the possibility for the electric field
from the ion to {\it not} lie in the $xz$-plane. This modification
is accomplished by replacing $E\cos\theta_{EB}$ with $E_z$,
$E\sin\theta_{EB}$ with $E_x+iE_y$ in $H_{16}$,
$H_{27}$, and $H_{38}$, and $E\sin\theta_{EB}$ with $E_x-iE_y$ in $H_{25}$,
$H_{36}$, and $H_{47}$; the matrix elements below the diagonal
are obtained from those above the diagonal by complex conjugation.

In what follows, $\vec{\psi}(t)$ is the 8-element vector holding the
amplitudes of each state, $\vec{r}(t)$ is the relative position
vector from the ion to the molecule, and $\vec{v}(t)$ is the
relative velocity vector.
We solve for both the classical and quantum dynamics using a
leapfrog algorithm. For the quantum state, knowing $\vec{\psi}(t)$
and $\vec{\psi}(t-\delta t)$ and knowing the $\vec{r}(t)$ the 
wave function at time $t+\delta t$ is found from
\begin{equation}
\vec{\psi}(t+\delta t)=\vec{\psi}(t -\delta t) - 2 i\delta t
\underline{H}(\vec{r}(t))\vec{\psi}(t)/\hbar
\end{equation}
where $\underline{H}(\vec{r}(t))$ is the $8\times 8$ matrix evaluated
for the E-field arising from the separation $\vec{r}(t)$. At this point
in the algorithm, the classical force can be calculated using the
Hellman-Feynman theorem:
\begin{equation}
\vec{F}(t) = -\langle\vec{\psi}(t)|
\vec{\nabla}\underline{H}(\vec{r}(t))|\vec{\psi}(t)\rangle .
\end{equation}
which will allow us to update the position and velocity vectors. Knowing
$\vec{v}(t-\delta t/2)$ and $\vec{r}(t)$, the position and velocity
at the next time step is computed from
\begin{eqnarray}
\vec{v}(t+\delta t/2)&=& \vec{v}(t-\delta t/2)+\vec{a}(t)\delta t\nonumber\\ 
\vec{r}(t+\delta t)&=&\vec{r}(t)+\vec{v}(t+\delta t/2)\delta t
\end{eqnarray}
where the acceleration $\vec{a}(t)=\vec{F}(t)/\mu$ and $\mu$ is the
reduced mass. These
steps are iterated, giving an algorithm that gives the motion of
the particles and time evolution of the internal states.

Since we assume the ion is much colder than the molecule, the
initial relative speed is the speed of the molecule. To compute
the cross section, we need to have a random set of initial positions
and impact parameters. We do {\it not} assume the initial velocity
is from a specific direction because there is an external
B-field which picks a specific form for the Hamiltonian. The direction
of the velocity vector is chosen from a uniform sampling on the
surface of a sphere.
The initial position is chosen to be
$\vec{r}(0)=\vec{b}-\hat{v}R$ where $R$ is a large initial distance
and $\vec{b}$ is a random point within a circle delimited by
a $b_{max}$ such that $\vec{b}\cdot\hat{v}=0$.
The initial conditions are such that we start the quantum state
in the highest energy eigenstate of $\underline{H}(\vec{r}(0))$.
The final time is chosen to be $2R/v$. For all parameters, we
test convergence with respect to $R$ (increasing $R$ until the
results don't change) and the $b_{max}$ (increasing the maximum
until the results don't change).

The physical parameters of interest are the inelastic and energy
transfer collision rates. The inelastic collision rate is found
from
\begin{equation}
\beta = v\pi b^2_{max}\langle 1 - P_8\rangle
\end{equation}
where $P_8 = |\langle \vec{\psi}_8|\vec{\psi} (t_f)|^2$ is the
probability the
molecule is still in the highest energy {\it eigenstate} at the final
time, $t_f$. The energy transfer collision rate is found from
\begin{equation}
\eta = v\pi b^2_{max} \langle KE_i-KE_f\rangle /KE_i
\end{equation}
where $KE_i$ is the initial kinetic energy of the molecule
and $KE_f$ is the final kinetic energy of the molecule. To find
the final kinetic energy of the molecule, we use center of mass
coordinates: $\vec{v}=\vec{v}_m-\vec{v}_i$ and $\vec{V}=
(M_m\vec{v}_m+M_i\vec{v}_i)/(M_m+M_i)$ where the $m,i$ subscripts
refer the molecule or ion respectively. The final velocity of the molecule is
found from $\vec{v}_{m,f}=\vec{V}+M_i\vec{v}(t_f)/(M_m+M_i)$ where
$\vec{v}(t_f)$ is the final relative velocity from the calculation
and the center of mass velocity $\vec{V}$ is a conserved quantity.
From these rates, we can solve for the rate of kinetic energy lost by the
molecule and the rate of population lost if we know the density of
ions. Taken as average quantities: $dP/dt = -n\beta P$ and
$dKE/dt=-n\eta KE$ where $n$ is the ion density,
$P$ is the population of trapped molecules and
$KE$ is the average kinetic energy.

\begin{figure}
\resizebox{160mm}{!}{\includegraphics{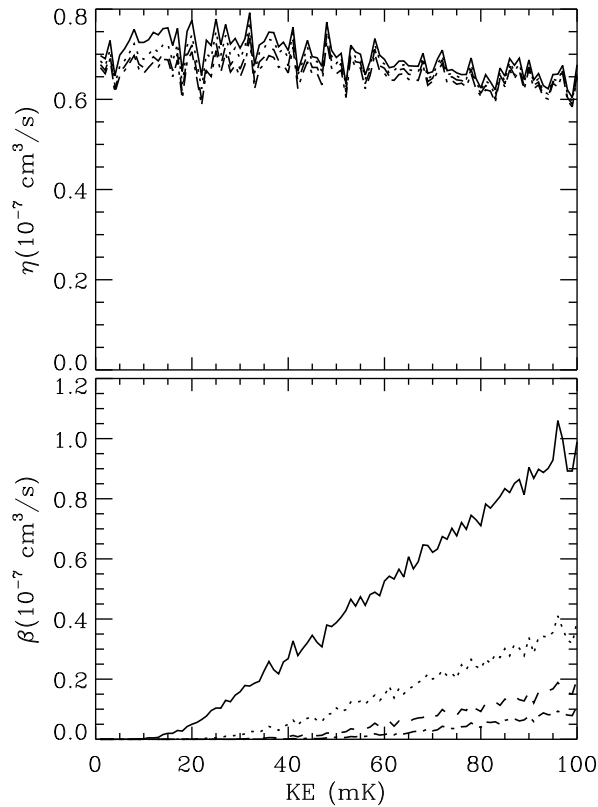}}
\caption{The energy loss rate, $\eta$, and population loss
rate, $\beta$, for neutral OH molecules colliding with
cold Mg$^+$ ions. The ions are assumed to be much colder
than the molecules and $KE$ is the kinetic energy of
the molecules. The OH are assumed to start in their highest
internal state in a B-field of 100~G (solid line),
200~G (dotted line), 300~G (dashed line), and 400~G (dash-dot line).}
\end{figure}

Figure~2 shows the rates as a function of OH kinetic energy
for the case where the ion is Mg$^+$.
For these parameters,
the energy loss rate is higher than the inelastic collision rate
over the whole energy range except for the 100~G case.
Another obvious feature is that the
inelastic collisions are more suppressed as the
B-field is increased, and
the inelastic rate has a threshold which is at larger $KE$
as the B-field increases. Both effects are
because the collisions become more
adiabatic as the splitting from the B-field increases.
Most importantly, the
energy loss rate, $\eta$, is approximately constant which results
from the fact that the repelling potential is approximately
$1/r^4$. For a classical Hamiltonian with a pure $1/r^4$ potential,
there is an exact scaling of any length
($L\propto 1/KE^{1/4}$) and speed ($v\propto KE^{1/2}$). Thus, classical rates,
$v L^2$, do not depend on $KE$. The smallest $KE$
calculated is 1~mK; see below for a discussion of the rates
at smaller $KE$.

\begin{figure}
\resizebox{80mm}{!}{\includegraphics{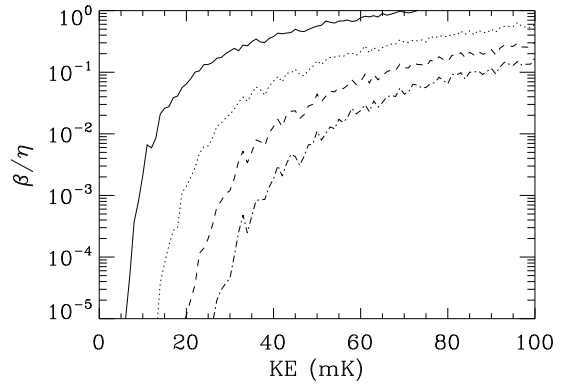}}
\caption{The ratio of parameters for Fig.~2; the line types are
the same as Fig.~2.}
\end{figure}

Figure~3 shows the ratio of population loss rate divided by
the energy loss rate for different B-fields.
From this plot, it's clear that the inelastic collisions
become completely unimportant as energy is removed from the molecules.
As an interesting point of comparison, the ratio is 0.1 for OH
kinetic energies of 22, 46, 66, and 90 mK for the 100, 200, 300,
and 400~G B-fields. The ratio is 0.01 for OH kinetic energies
of 13, 26, 39, and 51~mK. The kinetic energy at which the ratio
reaches a specific low value approximately scales with the
B-field.

\begin{figure}
\resizebox{160mm}{!}{\includegraphics{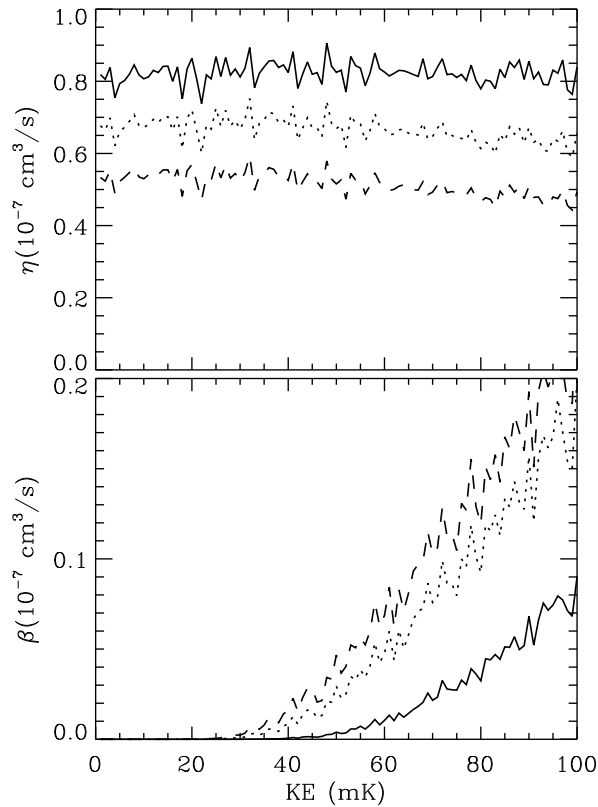}}
\caption{Similar to Fig.~2, except all lines are for 300~G. The
difference is the solid line is for OH collisions with $^9$Be$^+$,
$^{24}$Mg$^+$, and $^{40}$Ca$^+$. This shows lighter ions do a
better job cooling the OH molecules although the difference is
not qualitative.}
\end{figure}

Figure~4 shows the rates as a function of OH kinetic
energy for the case of 300~G for Be$^+$, Mg$^+$, and Ca$^+$.
There is the interesting trend that the inelastic collision
rate is smaller and the energy loss rate is larger as the
ion mass decreases. This trend is not surprising. For lighter
ions, the energy transfer
rate increases because it is easier to accelerate ion.
For lighter ions, the distance of closest approach is larger
(for the same reason) which suppresses the inelastic rate.

An important question is how many ions will be needed to produce
substantial cooling. {\it The limiting factor is the inelastic
collision between two molecules.} Reference~\cite{QB} (Fig.~4) has
an inelastic rate at 0~V/cm of $\sim 4\times 10^{-10}$~cm$^3$/s for
500~G and 50 mK, $\sim 1\times 10^{-11}$~cm$^3$/s for
500~G and 1 mK, and $\sim 1\times 10^{-11}$~cm$^3$/s for
500~G and 1 $\mu$K. These rates are much smaller than those in
Figs.~2 and 4 because the interaction distance is smaller for
the OH-OH collisions. If the molecules and ions are equally
distributed with an energy $\sim 50$~mK, then there
can be somewhat less than 1 ion per 200 molecules. Note that
once the molecule energy is $\sim 1$~mK or less, then there can be
approximately 1 ion per 10,000 molecules. In Ref.~\cite{SHY},
they estimate a peak density of OH of $\sim 5\times 10^{10}$~cm$^{-3}$
with $\sim 10^6$~molecules. For these parameters, sympathetic
cooling should be effective with less than 10,000 ions and,
perhaps, with as few as 100.

One important question is the energy dependence of the energy
loss rate, $\eta$, for energies below 1 mK. To get an idea,
we performed a quantum calculation for the energy loss rate
for a repelling $1/r^4$ potential. The actual potential depends
on both $r$ and $\cos\theta$ but we can obtain the trend with
respect to energy just from the isotropic interaction.
We fit the dependence
of the highest energy eigenstate as a function of E-field
and the angle with respect to the B-field. For low field
strengths, the energy can be written as
\begin{equation}
\epsilon_8 = \epsilon_8(E=0)+\frac{1}{2}(\alpha_0+\alpha_2 P_2
(\cos\theta ))E^2
\end{equation}
where $P_2$ is a Legendre polynomial,
$\alpha_0=7.5\times 10^{-10}$~mK/(V/m)$^2$, and
$\alpha_2=7.0\times 10^{-10}$~mK/(V/m)$^2$. For the quantum
scattering calculation, we set $\alpha_2=0$ and used
a potential $C_4/r^4$ with $C_4 = (\alpha_0/2)(e/4\pi\varepsilon_0)^2$.
The scattering could be treated using a method analogous to
Ref.~\cite{SWG}, but, instead, we numerically solved the
radial Schr\"odinger equation using a Numerov algorithm.

The quantum energy loss rate, $\eta$, changed by less
than 1\%
between 10~mK and 10~nK which is the behavior expected
from {\it classical} scaling. The energy loss rate
is
\begin{equation}
\eta\propto v\int_{-1}^1 (1-\cos\theta )\frac{d\sigma}{d\cos\theta}d\cos\theta
\end{equation}
with the proportionality constant independent of energy. As the energy
decreases, the cross section decreases but the differential cross
section is less strongly peaked at $\cos\theta =1$. The near constant
quantum rate for the isotropic potential suggests the energy loss rate
$\eta\sim 10^{-7}$cm$^3$/s for the full potential down to
$\sim 10$~nK.

We have not found a mechanism due to the ion collisions that would
prevent the molecules from reaching degeneracy. One possible
experimental arrangement could have the ions in a smaller spatial
region than the cold molecules at $\sim 10$'s~mK. The density
of molecules increases as the temperature decreases proportional
to $1/T^{3/2}$. In going from 50~mK to 1~mK, the OH density
increases by a factor of a few hundred but the inelastic collision rate
between two OH molecules also decreases by a factor of a few tens
which means the rate of losing molecules does not increase
proportional to the density.
An ion density $\sim 1/1,000$ that of the target
OH density would lead to OH temperatures substantially below 1~mK.
Thus, the space charge of the ions would be a minor perturbation.
After cooling, the
ions can be removed from the trap without disturbing the cold molecules.

We have performed calculations of collisions between an OH molecule
and a cold positive ion in a magnetic field. We find that the collision
is adiabatic at low temperature which means the inelastic cross
section is exponentially suppressed at low energy. We presented results for
100, 200, 300, and 400~G B-fields and found that the
cooling behavior is more favorable for larger fields.
The limitation to the cooling will be the inelastic collision between
pairs of OH molecules. Since this inelastic collision rate also decreases
rapidly with decreasing temperature, it seems likely that a small
fraction of cold, positive ions could sympathetically cool
neutral molecules. The lower bound on the OH temperature can
not be estimated from the data in this paper but we expect it
to be well below 1~mK. Because the mechanism that suppresses the
inelastic ion-molecule collision is a generic property of perturbed
quantum systems, the results in this paper should be generally
applicable for neutral molecules with magnetic and electric dipole
moments. Furthermore, since the inelastic molecule-molecule transition
is due to a shorter range interaction, it seems likely that cooling
will be possible for many, if not most, molecules with magnetic
and electric dipole moments.

We thank C.H.~Greene for pointing out relevant earlier work
and insightful conversations.
This work was supported by the Office of Basic Energy Sciences, U.S.
Department of Energy.

\end{document}